%% file: Livshits_et_al_2017_aps.tex
\documentclass[aps, twocolumn,showkeys,preprintnumbers,amsmath,amssymb]{revtex4}
\pdfoutput=1

\usepackage{graphicx}
\usepackage{epstopdf}
\usepackage{dcolumn}
\usepackage{bm}
\usepackage{pdflscape}
\usepackage{longtable}
\usepackage{threeparttablex}
\usepackage{multirow}
\usepackage{gensymb}
\usepackage{hyphenat}
\begin{document}

\preprint{}

\title{Catalog of Hard X-ray Solar Flares
Detected with Mars Odyssey/HEND\\
from the Mars Orbit in 2001-2016}

\author{M.A.~Livshits}
 \email{maliv@mail.ru}
\affiliation{
Pushkov Institute of Terrestrial Magnetism, Ionosphere and Radiowaves Propagation of Russian Academy of Sciences (IZMIRAN)
}%

\author{I.V.~Zimovets}
\affiliation{
Space Research Institute (IKI) of Russian Academy of Sciences, Moscow, Russia
}%

\author{D.V.~Golovin}
\affiliation{
Space Research Institute (IKI) of Russian Academy of Sciences, Moscow, Russia
}

\author{B.A.~Nizamov}
\affiliation{
Faculty of Physics of Lomonosov Moscow State University, Moscow, Russia
}
\affiliation{
Sternberg Astronomical Institute, Lomonosov Moscow State University, Moscow, Russia
}

\author{V.I.~Vybornov}
\affiliation{
Space Research Institute (IKI) of Russian Academy of Sciences, Moscow, Russia
}

\author{I.G.~Mitrofanov}
\affiliation{
Space Research Institute (IKI) of Russian Academy of Sciences, Moscow, Russia
}

\author{A.S.~Kozyrev}
\affiliation{
Space Research Institute (IKI) of Russian Academy of Sciences, Moscow, Russia
}

\author{M.L.~Litvak}
\affiliation{
Space Research Institute (IKI) of Russian Academy of Sciences, Moscow, Russia
}

\author{A.B.~Sanin}
\affiliation{
Space Research Institute (IKI) of Russian Academy of Sciences, Moscow, Russia
}

\author{V.I.~Tretyakov}
\affiliation{
Space Research Institute (IKI) of Russian Academy of Sciences, Moscow, Russia
}


\begin{abstract}
The study of nonstationary processes in the Sun is of great interest, and lately, multiwavelength observations and registration of magnetic fields are carried out by means of both ground-based telescopes and several specialized spacecraft (SC) on near-Earth orbits. However the acquisition of the new reliable information on their hard X-ray radiation remains demanded, in particular if the corresponding SC provide additional information, e.g. in regard to the flare observations from the directions other than the Sun-Earth direction. In this article we present a catalog of powerful solar flares registered by the High Energy Neutron Detector (HEND) device designed in the Space Research Institute (IKI) of Russian Academy of Sciences. HEND is mounted onboard the 2001 Mars Odyssey spacecraft. It worked successfully during the flight to Mars and currently operates in the near-Mars orbit. Besides neutrons, the HEND instrument is sensitive to the hard X-ray and gamma radiation. This radiation is registered by two scintillators: the outer one is sensitive to the photons above 40 keV and the inner one to the photons above 200 keV. The catalog was created with the new procedure of the data calibration. For most powerful 60 solar flares on the visible and on the far sides of the Sun (in respect to a terrestrial observer), we provide time profiles of flare radiation, summed over all the channels of X-ray and in some cases of gamma-ray bands as well as the spectra and characteristics of their power law approximation. We briefly discuss the results of the previous articles on the study of the Sun with HEND instrument and the potential of the further use of these data.
\end{abstract}

\maketitle

\section{\label{sec:level1}Introduction}
Development of nonstationary processes in the Sun is accompanied by substantial plasma heating and particle acceleration. These phenomena are actively investigated with both ground-based instruments and space astronomy methods. One of such nonstationary processes is a solar flare. They are observed in a wide range of electromagnetic radiation,  from radiowaves to gamma rays with progressively higher temporal and spatial resolution.

In the last decade one of the main instruments working in X-ray and gamma-ray energy bands is the RHESSI solar space observatory, which allows to obtain not only profiles and spectra of flares, but also their images in hard X-rays (HXR) and, in some cases, in gamma rays. The images of the Sun and solar flares in various bands of EUV and soft X-rays have been obtained onboard the spacecraft (SC) SOHO, CORONAS-F, STEREO, Coronas-Foton and GOES Imager as well. In recent years, a large amount of data on the radiation and magnetic fields in the Sun has been collected during the observations with Hinode and SDO.

In spite of the vast information on the characteristics of plasma and accelerated particles a systematic study of flares, especially their HXR radiation, is of great interest. In April 2001 the Mars Odyssey SC was launched towards Mars. Its equipment (namely, the gamma-ray spectrometer GRS \cite{Boynton04}) includes a detector of high energy neutrons (HEND). This device designed in Space Research Institute (IKI) of Russian Academy of Sciences is intended to map the neutron radiation of the Mars surface in order to study the hydrogen (water) abundance  in the planet's material \cite{Mitrofanov03, Sanin04}. Besides neutrons this device is also capable of detecting photons with energies from 40 keV to 2 MeV, therefore HEND data contain the records of such events as solar flares and cosmic gamma-ray bursts.

Continuous operation of the device during the interplanetary cruise and subsequently in the orbit around Mars allowed in many cases to obtain time profiles and X-ray spectra throughout the whole time of the development of the events, filling the gaps both in the ground-based observations and the observations from the near-Earth orbits. HEND detected mainly flares of M and X GOES classes. Even the profiles and spectra of the most powerful phenomena which could not be recorded by other SC were entirely registered. Measuring of HXR fluxes from different directions in some cases allowed to distinguish the contribution to the total flux from different parts of the source localized near the apex or near the footpoints of the loops and to clarify if the directivity of this radiation exists. Thereby the joint analysis of HEND data with the collection of data obtained with RHESSI, STEREO and other SC allows to investigate nonstationary processes in more detail. HEND has registered flares on the far side of the Sun with respect to a terrestrial observer and in a number of cases one can study a number of secondary processes.

HEND facility has been working stably since 2001, and there have been registered about 100 flares during 16 years. Some results have already been published in \cite{Livshits05, Kashapova08, Livshits11, Vybornov12}. The most interesting appeared to be a study of the phenomena observed on the solar disk from the near-Mars and near-Earth orbits (with RHESSI). An experience of investigating the flaring activity in the course of a full turnover of the Sun before the launch of STEREO was also interesting. In particular, we observed the rise of the group 10484 at the beginning of the period of very high flaring activity in late October 2003; the arrival of accelerated electrons and, subsequently, of a cloud of hot plasma \cite{Livshits11} from a behind-the-limb flare has been registered. This allowed to clarify the connection between gas-dynamic processes and particle acceleration, particularly, the emergence of a coronal mass ejection (CME) on the late stage of flares, to give an interpretation of some unusual phenomena in the interplanetary space and magnetosphere.

Mars Odyssey, as well as some other astrophysical SC such as Suzaku \cite{Suzaku}, was not intended to investigate the Sun. On both Mars Odyssey and Suzaku, in the data analysis one used a reference to other SC, particularly to RHESSI data. This is why some questions of solar investigations, e.g. a question on the directivity of HXR or accurate comparison of data from several SC, could not be solved completely. One of the problems of using such data has been the need of independent calibration of the instruments. Moreover, a question of the dependence of the sensitivity of the equipment on the orientation of the SC with respect to the Sun has not been investigated so far which led in some cases to the distortion of the measured HXR fluxes.

Below in this work we give a brief description of the equipment, present a new calibration of the X-ray measurements and the results of the study of their dependence on the SC orientation with respect to the Sun. Further, a catalog of events is described which contains time profiles of 60 events in X-ray and, in several cases, in gamma-ray bands, as well as the HXR spectra. We briefly discuss the earlier obtained results of the study of the Sun with HEND facility and some statistical relations for the quantities given in the catalog; comments concerning some of the events under discussion are given.

\section{On the HEND Facility}
High Energy Neutron Detector HEND is designed in the laboratory of space gamma spectroscopy of Space Research Institute (IKI) of Russian Academy of Sciences. It is a part of the gamma-ray spectrometer GRS. HEND includes two groups of detectors. The first one combines three proportional counters which use $^3$He isotope, the second one consists of two scintillators which register the hard X-ray and gamma radiation.

The scintillation block is implemented as two cylindrical scintillators, one inside the other. The outer scintillator is a CsI crystal, and the inner is a stilbene crystal. Apart from neutrons stilbene is capable of registering photons with energies in 60 keV -- 2 MeV band. However since this scintillator was located inside the other, the lower energy threshold was about 200 keV. The outer detector is also sensitive to the photons with energies above 40 keV. The energy band of each of these two detectors is subdivided to 16 channels which enables obtaining information on the spectrum of the registered radiation. The temporal resolution in each spectral channel is about 20 s. There is another type of data, namely time profiles with the resolution of 0.25 and 1 s for the inner and outer scintillators respectively. These time profiles are the count rates of the X-ray and gamma-ray radiation in all the channels of the inner and outer scintillators.

\section{On the calibration of the outer HEND detector}
The outer scintillation detector (OSD) of HEND has not been intentionally calibrated for the measurement of the HXR and gamma radiation before the launch of Mars Odyssey SC.

Moreover, the device does not contain calibration radioactive sources. For the conversion of the measurements from the ``count/s'' units to the physical units of the radiation flux density ``photon/s/cm$^2$/keV'', we calculated the response matrices of OSD by means of Geant4 \cite{Agostinelli03} and MCNPX \cite{mcnpx}. The calculations were carried out with no account of other HEND detectors, photomultipliers, electronics and the mounting board on the SC. Obviously, such an approach provides approximate results. However their accuracy is quite satisfactory. To make sure of this, we compared the energy spectra of HXR and gamma spectra obtained with OSD/HEND and RHESSI \cite{Lin02} for a number of solar flares observed with both instruments from close position angles.

\begin{figure}
\includegraphics[scale=0.5]{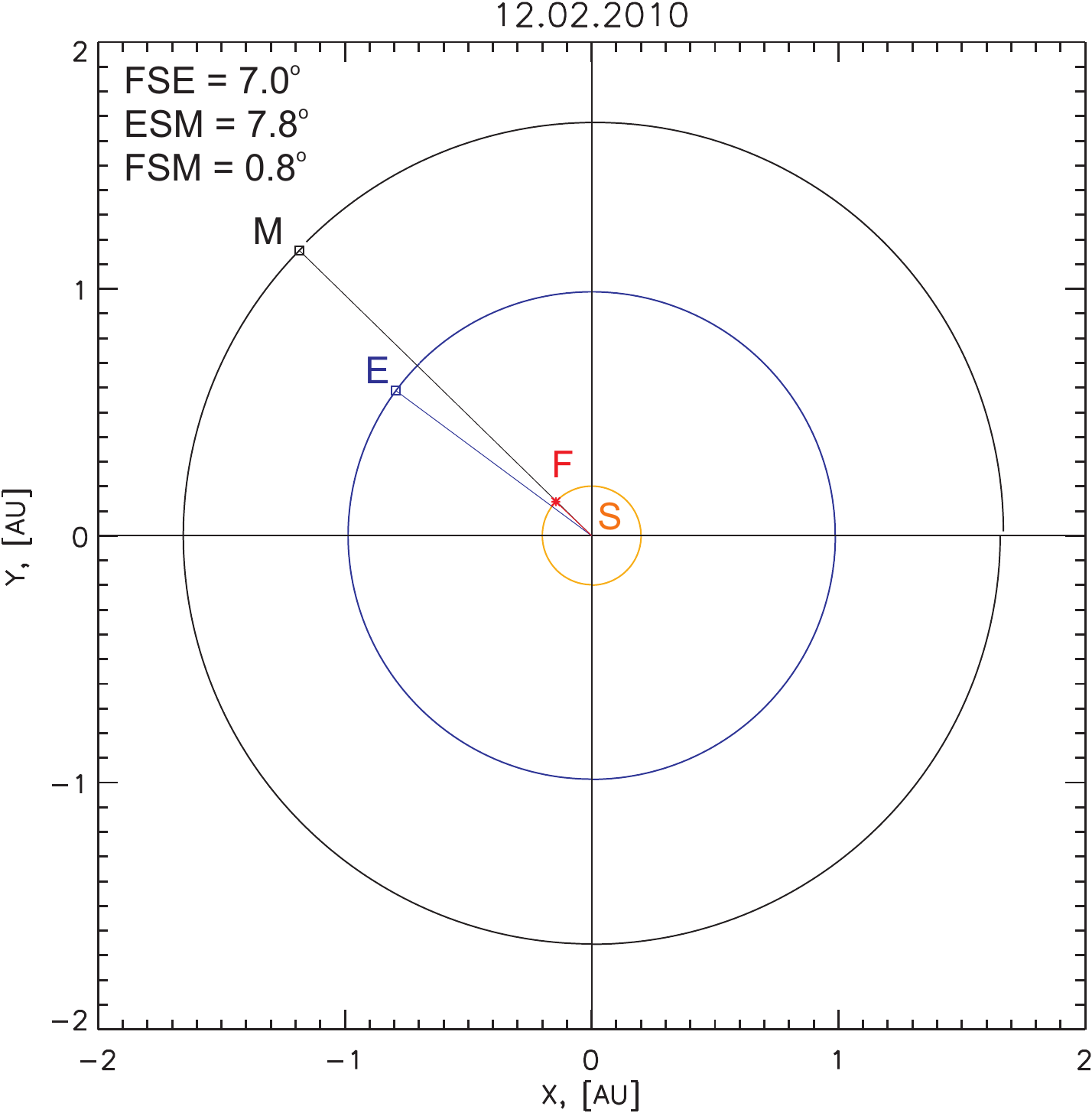} 
\caption{\label{orbits}The schematic illustration of the relative angular disposition of Earth (E), Mars (M), the Sun (S) and the flare (F) SOL2010-02-12T11:19 (not to scale).}
\end{figure}

\begin{figure}
\includegraphics[scale=0.5]{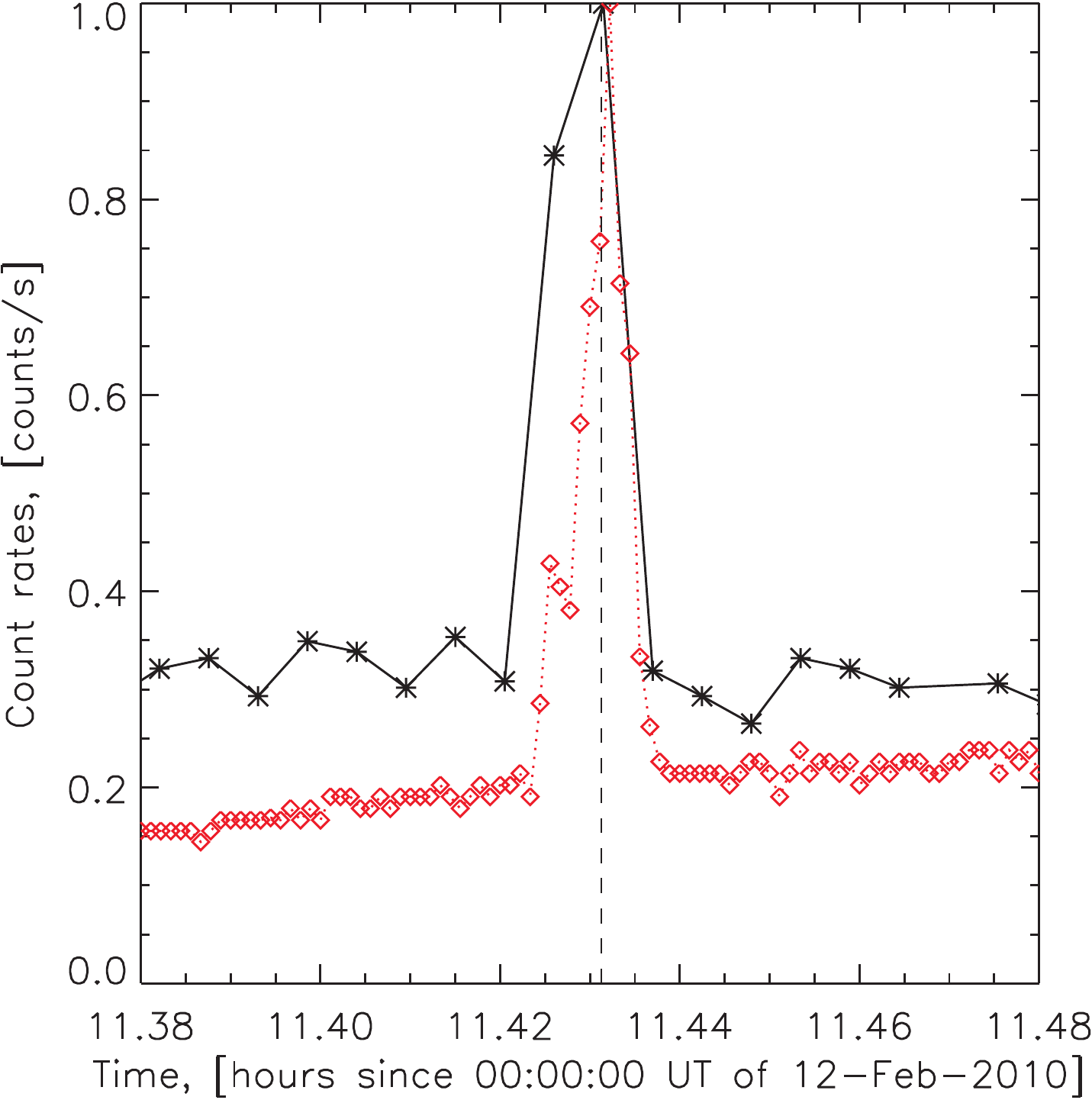} 
\caption{\label{prof_ivan}The count rates of the HXR radiation of the SOL2010-02-12T11:19 flare normalized to the maximum value according to the RHESSI data (red diamonds connected with a red dotted line) in the 50-100 keV channel with 4 s cadence and OSD/HEND (black asteriscs connected with a black solid line) in the channel 3 (86.6-107.7 keV) with 20 s cadence. The background is not subtracted.}
\end{figure}

As an example we describe the results of the comparison for an M8.3 flare which occurred on 12 Feb 2010 (SOL2010-02-12T11:19) near the center of the solar limb as observed from Earth (N26E07). In this case the ``Earth -- center of Sun -- flare'' angle was about 7\degree{}, the ``Mars -- center of Sun -- flare'' angle was about 1\degree{} and the ``Mars -- center of Sun -- Earth'' angle was about 8\degree{} (Fig.~\ref{orbits}). The time profiles of the count rates of HXR flare radiation measured with both instruments are given in Fig.~\ref{prof_ivan} with the correction for the difference of the distances from the Sun to Mars and to Earth. One sees that the profiles are very similar, however in the RHESSI profile (with the time step of 4 s) there is a small peak which is not seen in the OSD/HEND profile obtained with the 20 s cadence.

\begin{figure}
\includegraphics[scale=0.5]{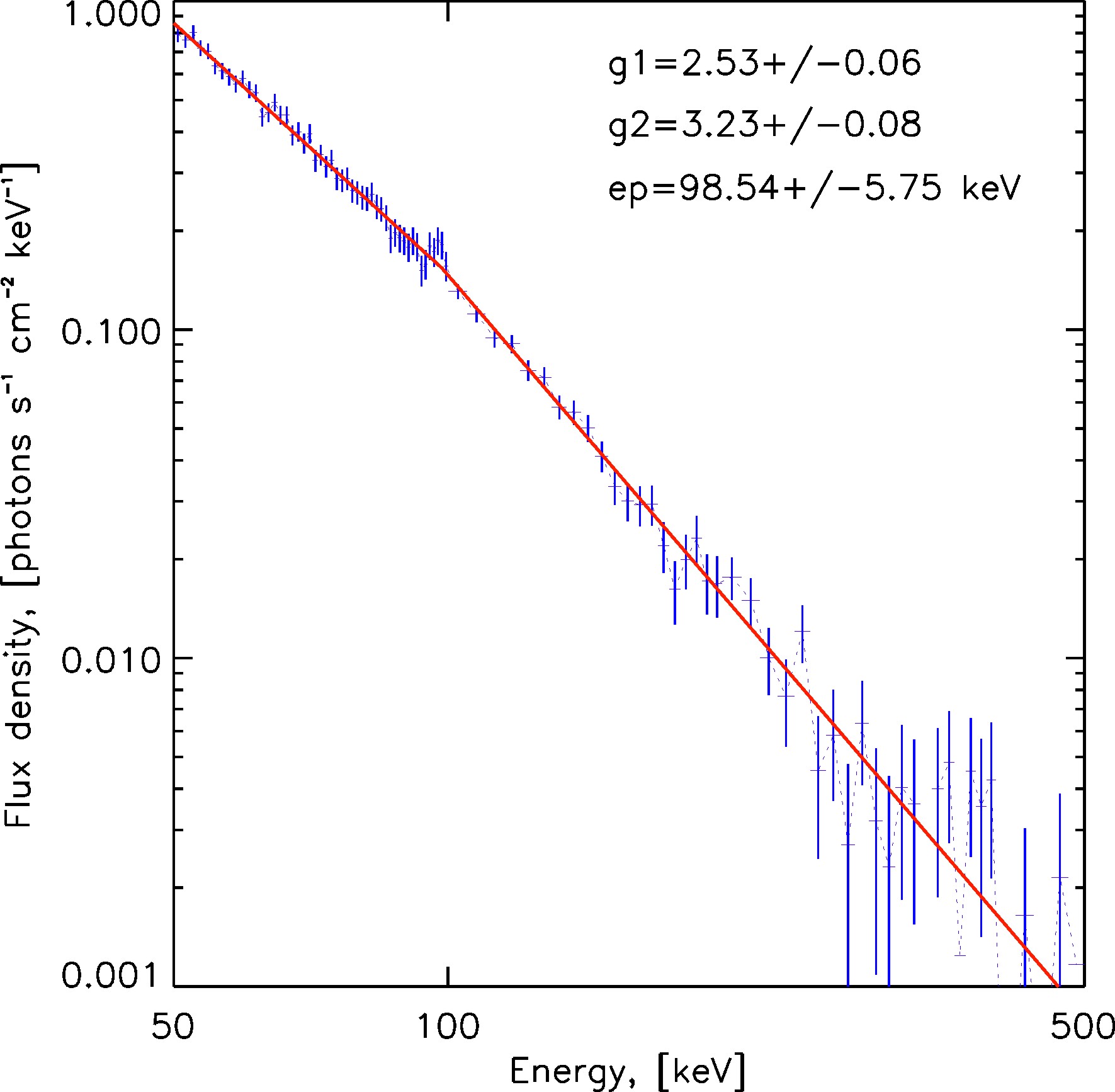} 
\caption{\label{rhessi}The spectrum of HXR radiation of the Sun with measurement errors in the vicinity of the peak (indicated by a vertical dashed line in Fig.~\ref{prof_ivan}) of the SOL2010-02-12T11:19 flare according to the RHESSI data with the background subtracted (blue vertical lines connected by a blue dashed line) and its least squares fit by a double power law function (red solid line) in the energy range 50-500 keV. The absolute values of the slopes (g1 and g2) and the break energy (ep) and their errors are indicated in the upper part of the figure.}
\end{figure}

In Fig.~\ref{rhessi} we give the energy spectrum of HXR radiation of the flare obtained from the RHESSI measurements in the 20-seconds time interval 11:25:52-11:26:12 UT which corresponds to the OSD/HEND measurement in the flare peak (the peak in Fig.\ref{prof_ivan}). The spectrum is obtained with a standard technique using the OSPEX package in SolarSoftWare \cite{Smith02}. In the energy range 50 -- 500 keV this spectrum (with the background subtracted) may be well approximated by a double power law function. In order to be compared with the HXR spectrum obtained with OSD/HEND, the RHESSI spectrum was integrated over the energy intervals corresponding to the energy channels of OSD/HEND. The integration is performed in two ways: 1) analytical integration of the fitted function; 2) bin-by-bin summation of the spectrum points. The results obtained with two methods are practically identical, however the first method allows (by means of interpolation) to obtain the spectra values in the energy ranges where the spectral flux values, according to RHESSI data, are formally below the background level, therefore in this work we only use the data obtained by analytical integration.

\begin{figure}
\includegraphics[scale=0.55]{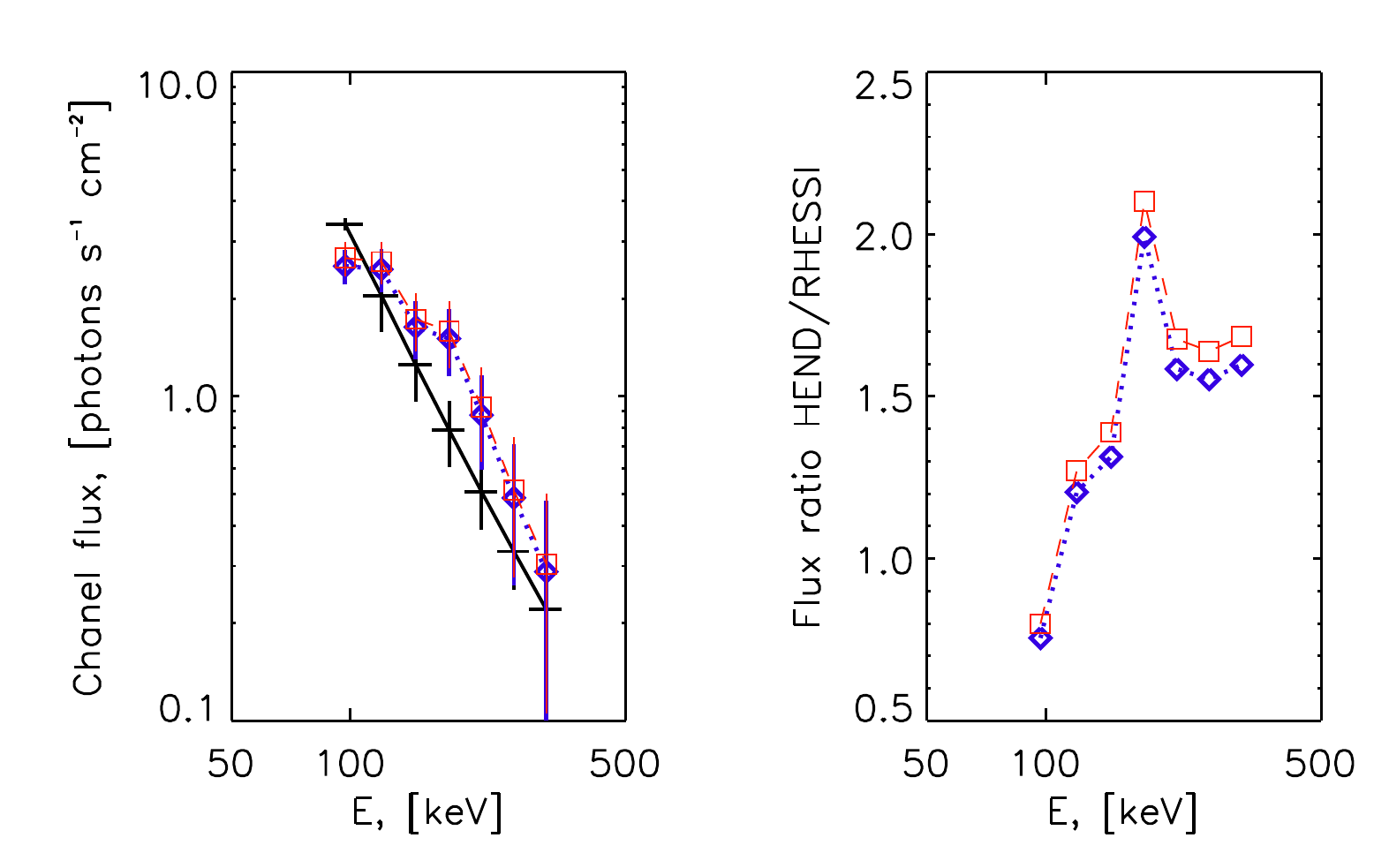} 
\caption{\label{comparison}\textit{Left panel.} Black bold crosses connected by a black solid line denote the HXR spectrum of the radiation in the peak of the SOL2010-02-12T11:19 flare (see Fig.~\ref{prof_ivan}) obtained via integration of the RHESSI data over the OSD/HEND energy channels (see Fig.~\ref{rhessi}). The measurement errors are shown with vertical dashes, the width of the energy channels --- with the horizontal lines. Blue diamonds connected by a blue dotted line and red squares connected by a red dash-dotted line denote the HXR spectra obtained from OSD/HEND in the flare peak using the response matrices calculated via modeling in Geant4 and MCNPX, respectively. \textit{Right panel.} The ratios of the spectra obtained from OSD/HEND and RHESSI shown in the left panel (the designations in both panels correspond to each other). One can see that the spectra obtained with the two instruments agree within a factor of 2.2. This indicates quite high precision of the OSD/HEND response matrices obtained in both Geant4 and MCNPX.}
\end{figure}

The results of comparison of the HXR spectra from OSD/HEND and RHESSI for the event discussed are given in Fig.~\ref{comparison}. The OSD/HEND spectrum shown in the left panel with diamonds and squares is obtained using the response matrices calculated, respectively, in Geant4 and MCNPX. One sees that the results obtained with Geant4 and MCNPX are practically identical (the difference is less than 5\%). It is also seen that the results obtained from the OSD/HEND and RHESSI data (crosses) are close to one another. The right panel of Fig.~\ref{comparison} shows that the spectra differ by factor of no more than 2.2. The ratio of the spectra is not equal for all the energy channels. The largest difference is observed in the OSD/HEND channel \#6 (161.056-195.079 keV). Nevertheless, we stress that as a whole one can see good agreement (in the limit of factor 2--3) of the spectra obtained with two different instruments. An analogous result is obtained for a number of other flares observed from close position angles. Thus, one can conclude that the response matrices obtained are of acceptable accuracy and may be used for the reconstruction of the HXR spectra from the OSD/HEND data. A more detailed description of the calibration of OSD/HEND will be given in the other paper.

\section{Event selection and data reduction}
\subsection{Event selection}
It is known that solar flares with powerful radiation in the range above 300--400 keV occur relatively rarely. During all the operation time HEND has registered only 3 such powerful events for which the time profiles in several energy channels of the inner scintillator were obtained. These are the flares on 25 Aug 2001, 27 Oct 2002 and 28 Oct 2003. The first and the last ones are well known due to the observations on Yohkoh and RHESSI. The flare on 27 Oct 2002 has occurred far behind the limb for a terrestrial observer. Therefore the data on the electromagnetic radiation of the solar flares are mainly from the outer detector and everything considered in the present paper refers to this detector unless otherwise pointed. The event selection was  made by the registration of the count rate in the integral channel.

The events with 3$\sigma$ excess above the background level (according to the registration by HEND with 20 s temporal resolution) were compared with the times of the flare maxima from the IZMIRAN catalog \cite{IZMIRAN} which is based mainly on the GOES data. Difficulties arose in rare cases, e.g. in September of 2011, when a bulk of particles from powerful CMEs and solar cosmic rays (SCR) reached Mars, and the radiation environment around the SC did not allow to register the HXR emission from flares in a time interval from several hours to 1--3 days. In some cases for more accurate identification, the RHESSI data were used since the operating range of this instrument is analogous to that of HEND. However since late 2001 Mars Odyssey had been in the near-Mars orbit and for this reason a part of the events could not be identified by the method described above due to the event seeing conditions of two SC. We therefore inspected the whole bulk of HEND data for events not identified from Earth. On finding such an event the decision whether it is a solar flare was made on the basis of several criteria. The first criterion is whether a CME or a type II radioburst was registered close in time to the event in question. Most often a CME was observed, in many cases marked in the LASCO database as a ``halo with no identified source'' which gave a ground to consider the event in the HEND data a solar flare. If a CME was absent, the time profile with 0.25 s cadence was investigated since in these profiles weak events were identified best of all. The spectrum was estimated as well. In some cases cosmic gamma-ray burst databases were checked in order for such events not to fall into the catalog. In recent years, a number of events are identified with the flares on the far side of the Sun registered by  STEREO.

\subsection{Fluxes and photon spectra}
We use here the spectral data of OSD/HEND,  sums of counts in a given energy channel over the integration time of about 19.75 s. Let us briefly describe the procedure of the conversion of the count rates to the physical fluxes (photon/cm$^2$/s/keV) at the distance of Earth orbit and the construction of the flare spectra.

OSD/HEND has 16 energy channels (they are numbered from zero) of which we use for the analysis those with numbers from 3 to 14. For the first three channels (0--2) energy boundaries are not precisely known. The upper boundary is not clearly defined for the last channel (15) as well. It registers photons with energies higher than the threshold which is equal to the energy of the upper boundary of the channel 14. The energy boundaries of the channels we used are provided in Table~\ref{Tab1}. Note that in the Mars Odyssey main program observations the outer detector serves only as an element of the anti-coincidence scheme and is not used for actual measurements. This does not require precise knowledge of the boundaries of softer channels and in the table we give only their approximate values.

\begin{table}
\caption{\label{Tab1}Energy channel boundaries of OSD/HEND.}
\begin{ruledtabular}
\begin{tabular}{c|c|c}
Channel & Lower boundary, keV & Upper boundary, keV \\
\hline
1 & $\approx 45$ & $\approx 65$ \\
2 & $\approx 65$ & 86.5 \\
3 & 86.568 & 107.683 \\
4 & 107.683 & 132.256 \\
5 & 132.256 & 161.056 \\
6 & 161.056 & 195.079 \\
7 & 195.079 & 235.632 \\
8 & 235.632 & 284.493 \\
9 & 284.493 & 344.09 \\
10 & 344.09 & 417.893 \\
11 & 417.893 & 510.993 \\
12 & 510.993 & 631.240 \\
13 & 631.240 & 791.506 \\
14 & 791.506 & 1014.72 \\[1mm]
\end{tabular}
\end{ruledtabular}
\end{table}

At the first step of the reduction the background was subtracted from the data. The background in the spectral data varied only slightly on the time scales comparable with the duration of a flare, except for several short periods of time associated with the presence near Mars of particles accelerated in solar flares and, in rare cases, with the particularities of the Mars investigation program. The background remained practically constant in the soft channels with the increasing of modulation towards harder ones. The period of this modulation is equal to half the orbital period of the SC.

\begin{figure}
\includegraphics[scale=0.5]{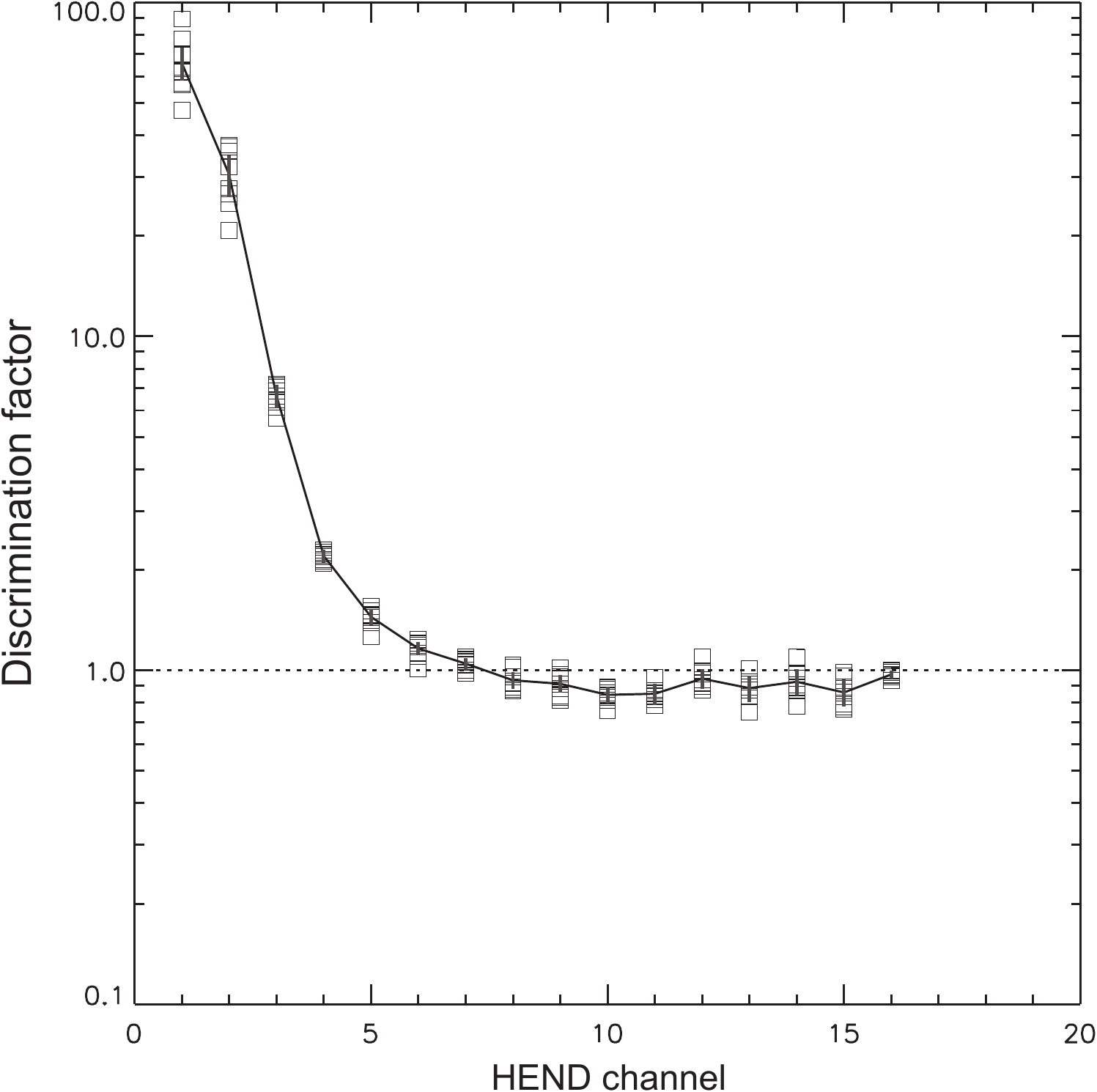}
\caption{\label{discriminator}The discrimination coefficients for various energy channels of OSD/HEND with the measurement errors (vertical dashes).}
\end{figure}

To account for the background the signal of the detector in each channel before the flare (sometimes after the flare) was approximated by a straight line or a parabola. To construct a spectrum one has to select a certain time instant. We chose the moment when the count rate in the channel 4 (counting from zero) reaches the maximum. Thus, for each flare we obtained a vector $\mathbf{I}_\mathrm{c}$ (``c'' stands for counts) each of twelve coordinates of which is the detector's count rate in one of the channels from 3 to 14. The corresponding vector of physical fluxes $\mathbf{I}_\mathrm{phot}$ is bound to it via a linear relation $\mathbf{I}_\mathrm{c} = A \mathbf{I}_\mathrm{phot} S d \Delta E$.  Here $A$ is the  response matrix of the detector. This matrix depends on the direction from which the photons fall onto the detector, to be more accurate, on the angle between the direction to the source and the symmetry axis of the detector. In the process of calibration matrices for five directions from 0 to 180\degree{} have been calculated. For intermediate directions the matrices were found via elementwise interpolation. The orientation of the detector with respect to the Sun at the moment of each flare was calculated with the SPICE package \cite{SPICE}. Further, $S$ is the cross-section of the detector as observed from a given direction, $d$ is the discriminator coefficient which is connected with the device operation features. The values of this coefficient with the errors for different channels are given in Fig.~\ref{discriminator}. Finally, $\Delta E$ is the width of a given energy channel. The knowledge of the count rate vector allows to find the vector of physical fluxes. In order to estimate the error of the physical flux we took into account the dispersion of the background flux, the error of the discriminator and the fact that the arrival of (flare) photons is a Poisson random process with a characteristic dispersion.

\begin{figure*}
\begin{minipage}{0.45\linewidth}
\includegraphics[scale=0.5]{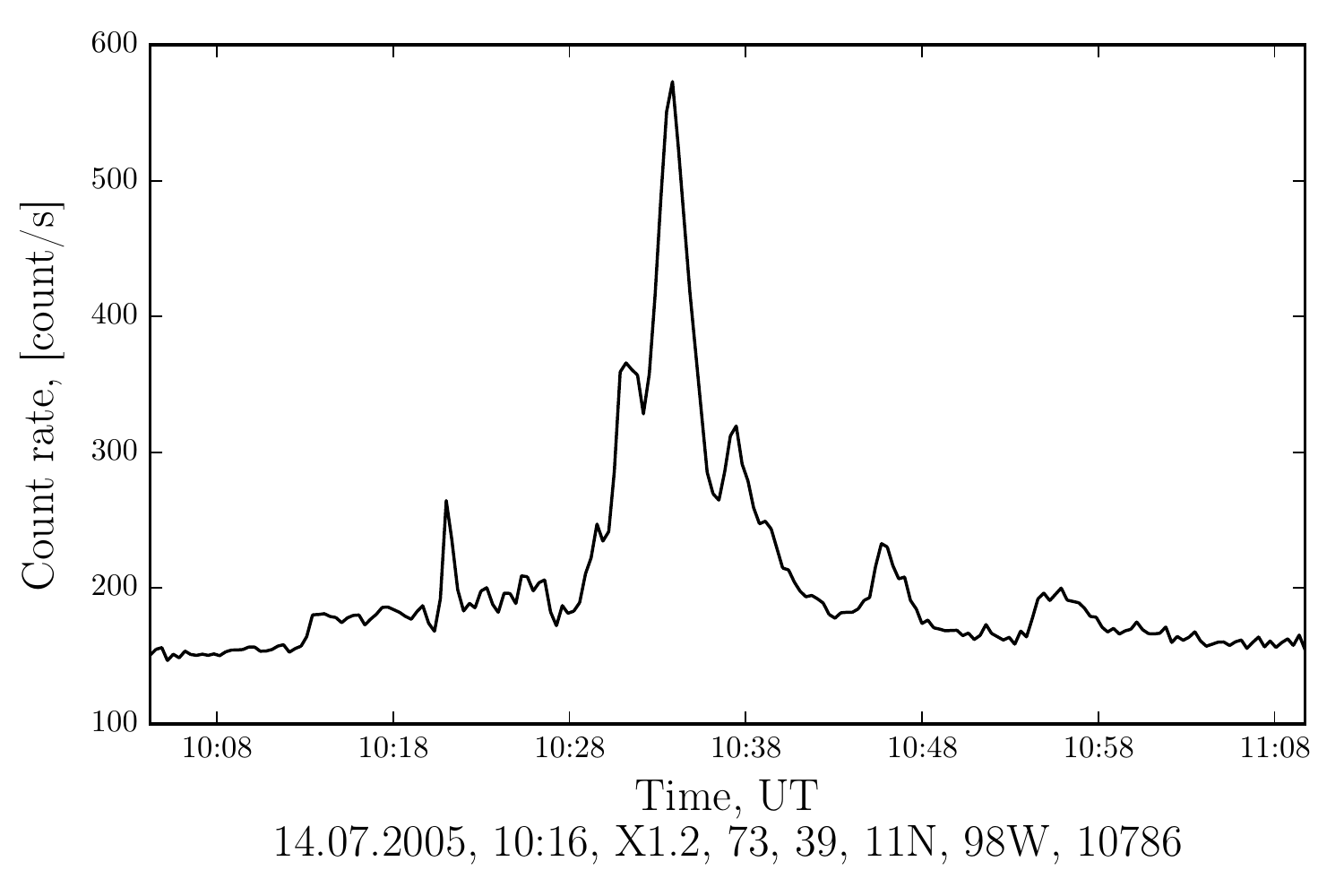}
\end{minipage}
\begin{minipage}{0.45\linewidth}
\includegraphics[scale=0.50]{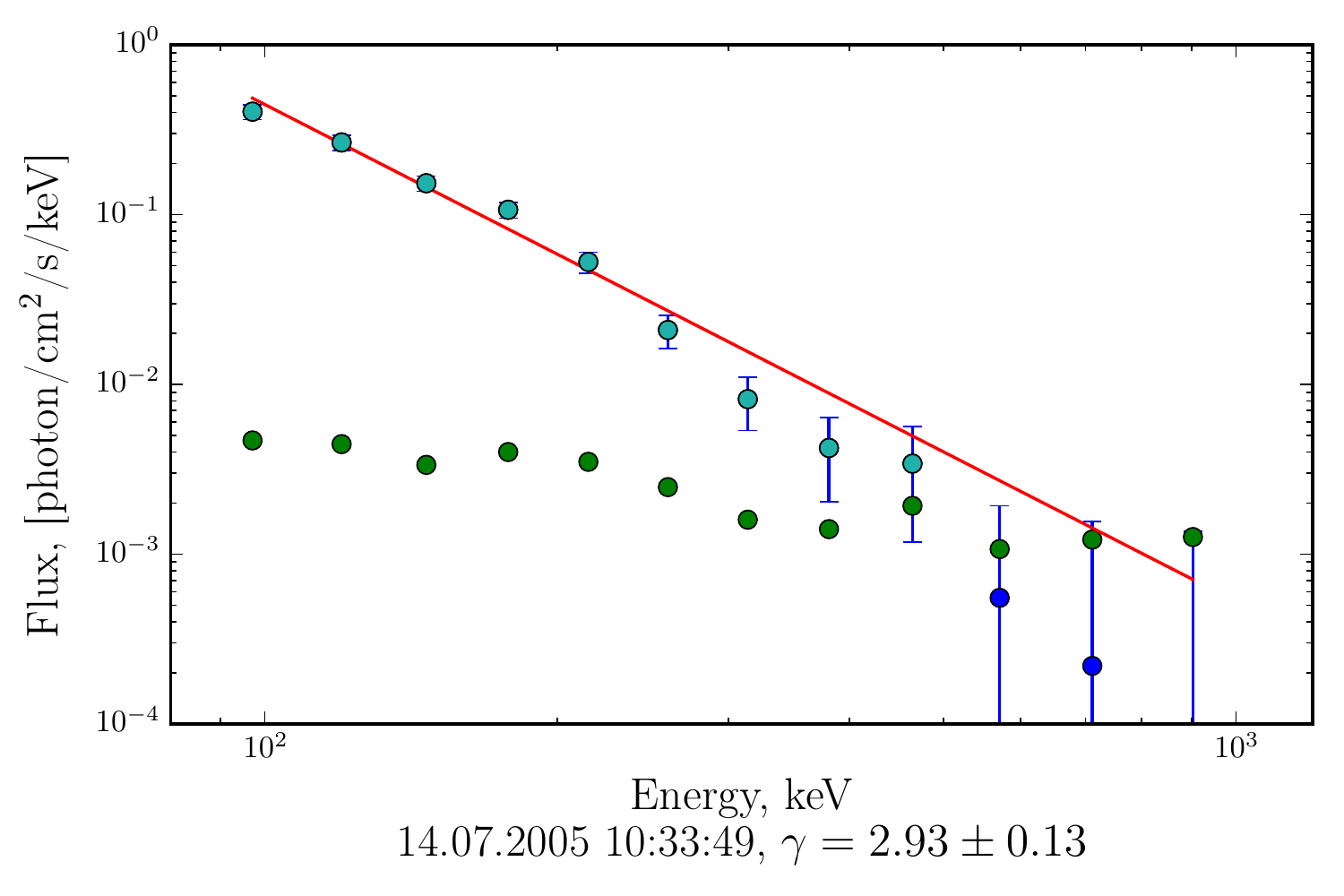}
\end{minipage}
\caption{\label{flare_lc}\textit{Left panel.} An example of the flare profile from the catalog. The label indicates that the event happened on 14 Jul 2005, the start time 10:16, the GOES X-ray class is X1.2, the total duration is 73 min, the rise time is 39 min, the active region coordinates are N11W90, the active region id is 10786 (all the times are given according to GOES). The ordinate values are the count rates in the integral channel of the outer HEND detector, the cadence is 19.75 s. \textit{Right panel.} The spectrum of this event. Green circles denote the quantity $3\sigma$ where $\sigma$ is the standard (root mean square) deviation of the background values; cyan circles denote the flare flux with the background subtracted in the channels where it exceeds $3\sigma$, blue circles denote the flux in the rest of the channels. The spectral fit (red line) is done only using cyan circles. The time in the label is according to HEND.}
\end{figure*}

Then the counts in the spectral channels used should be approximated by a model spectral curve. As such a model we used a power law, i.e. $I(E) = aE^{-\gamma}$. For model parameters estimation we only used the fluxes in the channels where the flux exceeds the background by no less than $3\sigma$, where $\sigma$ is the root mean square deviation of the background. The model parameters obtained along with the errors for all the investigated flares are given in the consolidated Table~\ref{Tab2}. The full description of the table is given in section~\ref{catalog}.

\section{Catalog of powerful flares} \label{catalog}
The catalog of powerful flares is presented as online supplement. It contains  the total count rate for each of 60 events in all the channels of the outer scintillator in pulses (counts) per second at the location of Mars Odyssey. The background subtraction was not performed. For a number of the most powerful events the gamma-ray profiles are provided, i.e. the sums over all the channels of the inner scintillator, again without the background subtraction. The following information is provided in the figure labels: the date, the GOES start time, GOES X-ray class, the full duration in minutes according to GOES, the rise time in minutes according to GOES, the active region coordinates where the flare took place, the active region id, and if then an optional index ``i'' is given, it means that the given profile is from the inner scintillator. As an example we give below, in section ~\ref{discussion}, a figure for the event on 14 Jul 2005 10:33 (Fig.~\ref{flare_lc}). In addition to the time profiles, the spectra of the investigated events are shown with the power law approximation in the energy range above 100 keV. The ordinate axis shows the fluxes near Earth in units ``photon/cm$^2$/s/keV'', the abscissa axis shows the energy in keV. Note that for weak soft X-ray events marked with an asterisk in the first column of Table~\ref{Tab2} HXR fluxes did not allow to fit a model with sufficient accuracy (see the discussion and Fig.~\ref{goes_gamma}). Moreover, when using the catalog, one has to keep in mind that during some events HEND was shaded by the body of the SC which inevitably lowers the reliability of the flare spectrum measurement. The results of the observations are compiled in the consolidated Table~\ref{Tab2}. Let us give the description of all its columns.

1. The index number of the event. An asterisk means that the HXR flux was too low to derive the spectrum parameters sufficiently  accurately.

2. The date.

3. UT time of the burst maximum in channel 4 (108 -- 132 keV) reduced to the distance from the Sun to Earth, i.e. the time of the event for a terrestrial observer.

4. Time in seconds which takes the electromagnetic radiation to cover the difference in distances from the Sun to Mars Odyssey and from the Sun to Earth. This was computed using SPICE package \cite{SPICE}.

5. GOES X-ray class.

6. The NOAA active region id. Note that for two events the GOES class is absent, but the active region is known. These events occurred on the far side of the Sun, but close to the limb. They are ascribed to that active region (which rose soon) in which CMEs and radiobursts associated with it took place at that time.

7. The flux in channel 3 (87 -- 108 keV) and its error. This is the flux at the time for which the flare spectrum was calculated. The flux at Earth is given.

8, 9. The parameters of the power law function $I(100)(E/100)^{-\gamma}$ fitting the radiation spectrum, where the photon energy $E$ is in keV and $I(100)$ is the estimate of the photon flux density at the energy of 100 keV.

10. The duration of the burst in channel 3. As the start of a flare was considered the first instant when the flux exceeded the background level by more than $2\sigma$. The end of the flare is the last such instant. Since we used the data with the cadence of 20 s, the durations are determined with the same accuracy.

11. The total number of counts in channel 3 in the course of the flare (fluence). This quantity was calculated as an integral of the count rate curve in channel 3 (without multiplication by the discriminator) multiplied by 20 since the data themselves are the count rates averaged over 20 second intervals. The sums of counts are also reduced to the distance to Earth (i.e. multiplied by $(r_\mathrm{M}/r_\mathrm{E})^2$ where $r_\mathrm{M}$ is the distance from the Sun to Mars Odyssey and $r_\mathrm{E}$  is the distance from the Sun to Earth).

12. The index showing whether there was an occultation of the detector by the body of the SC. ``No'' (``0'' in the table) means that the solar disk was visible from the origin of the reference frame associated with HEND. ``Yes'' (``1'' respectively) means that the solar disk is covered by the body of the SC and the measurements, especially in the soft spectral region, could be distorted. The information on the occultation of HEND by the SC body was obtained from the data of SPICE system and an intentionally designed simplified model of the SC.

\begin{ThreePartTable}
  \begin{TableNotes}\footnotesize
\item \textbf{Note.} The following is given in the head: the event id \#, the asterisc means that the HXR flux was too low to determine the spectral parameters sufficiently accurately; the date; the UT time of the burst maximum in channel 4 (108--132 keV) for a terrestrial observer; $\delta t$ --- the time in seconds which takes the electromagnetic radiation to cover the difference in distances from the Sun to the SC and from the Sun to Earth; the GOES X-ray class; NOAA --- the active region id; F3 --- the flux in channel 3 (87--108 keV) in phot/cm$^2$/s/keV units; the parameters of the power law function $I(100)(E/100)^{-\gamma}$ fitting the radiation spectrum; $\Delta t$ --- the duration of the burst in channel 3; $\Sigma$(E3) --- the total number of counts in channel 3; U --- indicates whether there was an occultation of the device by the SC (``1'' --- was, ``0'' --- was not).
  \end{TableNotes}
\begin{longtable*}{p{0.5cm}|p{1.8cm}|p{1.4cm}|p{0.9cm}|p{1cm}|p{1.1cm}|p{2.5cm}|p{2.3cm}|p{1.7cm}|p{0.8cm}|p{1.5cm}|p{0.5cm}}
\caption{\label{Tab2}Catalog of powerful solar flares.}\\
\hline
\hline
\multirow{2}{*}{\#} &
\multirow{2}{*}{Date} &
\multirow{2}{*}{Time} &
\multirow{2}{*}{\parbox[t]{1cm}{$\delta t$, s}} &
\multirow{2}{*}{\parbox[t]{1cm}{GOES class}} &
\multirow{2}{*}{\parbox[t]{1cm}{NOAA}} &
\multirow{2}{*}{\parbox[t]{1.7cm}{F3, $10^{-2}$}} &
\multicolumn{2}{|c|}{$F = I(100) (E/100)^{-\gamma}$} &
\multirow{2}{*}{$\Delta t$, s} &
\multirow{2}{*}{\parbox[t]{2cm}{$\Sigma$(E3)}} &
\multirow{2}{*}{U}\\\cline{8-9}
  &  &  &  &  &  &  & $I(100), \times 10^{-2}$ & $\gamma$ &  &  &\\
\hline\endfirsthead
\caption*{\textbf{Table~\ref{Tab2}.} Continued.}\\
\hline
\hline
\multirow{2}{*}{\#} &
\multirow{2}{*}{Date} &
\multirow{2}{*}{Time} &
\multirow{2}{*}{\parbox[t]{1cm}{$\delta t$, s}} &
\multirow{2}{*}{\parbox[t]{1cm}{GOES class}} &
\multirow{2}{*}{\parbox[t]{1cm}{NOAA}} &
\multirow{2}{*}{\parbox[t]{1.7cm}{F3, $10^{-2}$}} &
\multicolumn{2}{|c|}{$F = I(100) (E/100)^{-\gamma}$} &
\multirow{2}{*}{$\Delta t$, s} &
\multirow{2}{*}{\parbox[t]{2cm}{$\Sigma$(E3)}} &
\multirow{2}{*}{U}\\\cline{8-9}
  &  &  &  &  &  &  & $I(100), \times 10^{-2}$ & $\gamma$ &  &  &\\
\hline\endhead
\input{table866.dat}
\hline
\hline
\insertTableNotes
\end{longtable*}
\end{ThreePartTable}

\section{Discussion} \label{discussion}
The characteristics of the HXR spectra from HEND data have been previously obtained with the use of the cross-calibration with RHESSI. This calibration is briefly described in the Appendix. The independent calibration which has now been performed allows to estimate the accuracy of the spectra characteristics, in particular, those obtained earlier with the recalibration Mars Odyssey/RHESSI. As a rule, for most of individual events new and old determinations of 100 keV photon fluxes differ by a factor of no more than 1.5 -- 2. The difference is most prominent for the weakest flares. Moreover, they are pronounced stronger when the flare was located near the limb for at least one instrument since in such observations the dependence of the flux on the location of the flare with respect to the limb and on the SC orientation which was not considered earlier is manifested  stronger.

\begin{figure*}
\begin{minipage}{0.45\linewidth}
\includegraphics{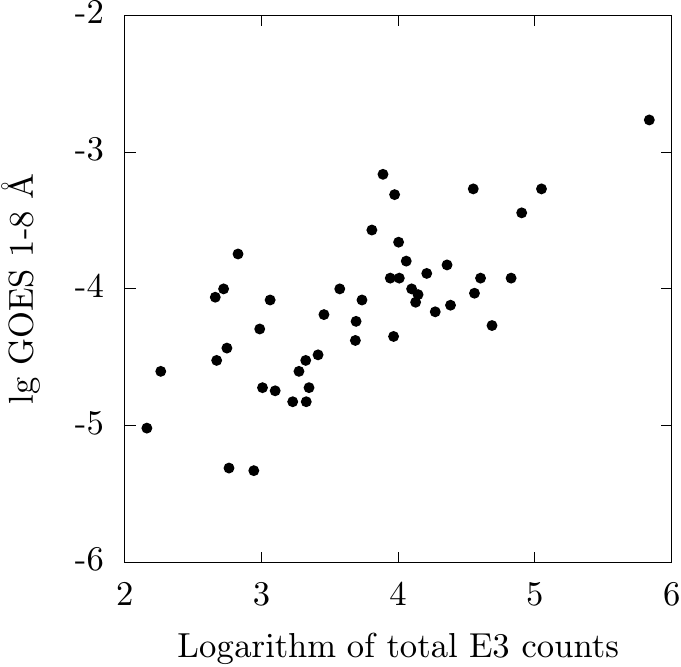}
\end{minipage}
\begin{minipage}{0.45\linewidth}
\includegraphics{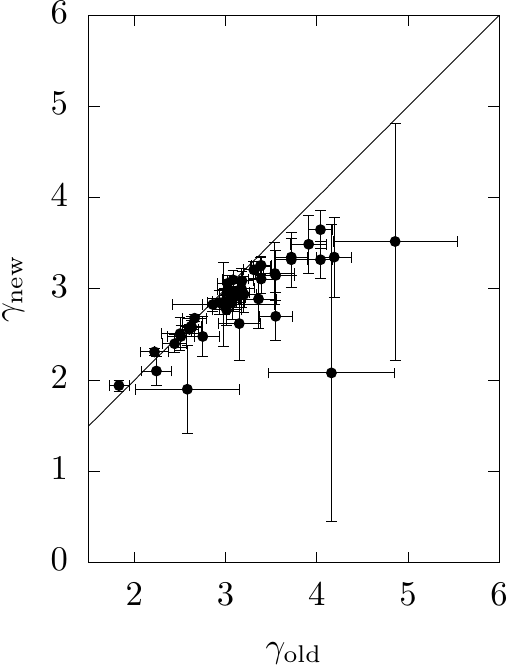}
\end{minipage}
\caption{\label{goes_gamma}\textit{Left panel.} The sum of HEND counts in channel E3 (87-108 keV) with the background subtracted vs. the power of the flares in soft X-rays. The number of counts is reduced to the distance to Earth. \textit{Right panel.} The power law slopes obtained with the old and new calibrations. The straight line corresponds to the exact equality ($\gamma_\mathrm{new} = \gamma_\mathrm{old}$).}
\end{figure*}

An example of the observations of the event on 14 Jul 2005 10:33 UT is given in the left panel of Fig.~\ref{flare_lc} with the count rate typical of M5 -- X1 flares. Here a precursor is manifested the registration of which on Mars Odyssey is partly distorted by sunrise effects at 10:14, as well as the main impulse with the count rate of several hundreds counts per second and additional maxima associated with the development of post-eruptive events (for more details on the phenomena on the Sun and the influence on the magnetosphere of this event on 14 Jul 2005 see \cite{Livshits11, Livshits_KI}). In the right panel of Fig.~\ref{flare_lc} the spectrum of the flare is given. The radiation of all the events in the range 87 -- 108 keV (channel 3) correlates well with the power of soft X-ray radiation in the maximum of the flare. This is illustrated by the left panel of Fig.~\ref{goes_gamma}, where the corresponding fluences and GOES 1--8\AA{} fluxes are plotted.

Besides the comparison of the temporal profiles for individual events, the comparison of spectral slopes appeared to be more vital for estimation of the reliability of the earlier data. In the right panel of Fig.~\ref{goes_gamma}, we plot the spectral slopes for old and new determinations. One sees two cases (12 May 2001, 04 Jun 2001) with large errors and harder slopes according to new determinations. The slope errors for these events are more than 30\%, we explain this by the high noise level. However the correlation coefficient after excluding these events is quite high ($r \approx 0.9$).

Thus, the first conclusion to be derived from our consideration is that with the accuracy stated above HEND data allow to determine the characteristics of HXR emission of solar flares. For most of heliophysical studies, such accuracy of fluxes and spectra determination appears to be sufficient. Therefore the main conclusions of previous works remain valid. In other words, the data of HEND and RHESSI agree well. One should take into account that the power law slopes refer to rather narrow spectral range where the spectral fitting was done. Therefore, the slope values we provide do not describe the character of the entire spectrum, in particular, they do not give any hint whether there is a break in the spectrum. The double power law can be revealed if the event is reliably detected in hard channels of the inner scintillator and is pronounced in gamma rays, at energies above 300 keV. Note also that in the catalog we give the time profiles according to the spectral data.

This conclusion is important not only for the HEND detectors, but also for analogous devices mounted on other SC not designed specifically for the observations of the Sun. For example, on the Suzaku SC \cite{Suzaku_catalog} recalibration of the solar measurements with the use of RHESSI data was performed and now one may use these data on the solar flares of various power more confidently.

However, in order to investigate more subtle effects on the basis of small intensity variations, e.g. the directivity of the HXR emission, one should have of course an independent calibration of the data.

Note also that in recent years HEND data may be used both for analysis of powerful events and for comparison with other instruments registering the solar radiation from the Earth -- Sun direction as well as from various directions, like STEREO A, B.

In the supplement to the electronic version of the article we provide the profiles of all the events in the integral channel of the spectral data (20 s cadence) of the outer detector and in some cases of the inner detector. The spectra and their power law fits at energies higher than 100 keV are also given. Some remarks on the individual events and corresponding references are collected. Besides, the previous data based on the cross-calibration with RHESSI are given along with the spectral fits starting from 50 keV.

\begin{acknowledgments}
The authors are thankful to W. Boynton, C. Shinohara, D. Hamara and other collaboration members who have been facilitating the operation of Mars Odyssey spacecraft since 2001 up to now. We intensively used the data on solar flares obtained with RHESSI. We are grateful to the RHESSI, SOHO and GOES collaborations for the opportunity to use the data in open access. SOHO is a cooperative project of the European Space Agency and NASA. The authors thank I.V. Kudinov for the help with the calculations in Geant4. This work is supported by the grant of Russian Foundation for Basic Research 14-02-00922 and partially by the 1.7P program of the RAS Presidium. The authors thank I.M. Chertok and A.V. Belov for fruitful discussions.
\end{acknowledgments}

\appendix*

\section{}

In order to facilitate the use of the time profiles given in the catalog and to estimate the accuracy of the values used in previous works on solar flares, we provide here the results of our previous analysis of a part of these data (consequently the list of the events has been extended). Earlier a procedure of cross-calibration --- comparison of the HEND count rates with the absolute measurements on the other SC (RHESSI) was used. For such a procedure, a number of solar flares were selected the radiation of which was registered by both SC. In order to ensure the observation of the same parts of the source, we excluded the events  that happened close to the limb  as observed from at least one SC. In order  for the possible directivity of the X-ray radiation not to distort the results of the fluxes comparison, we considered the phenomena for which the angle between the directions to the flare from RHESSI and Mars Odyssey did not exceed 60\degree. It is known that RHESSI did not register the radiation flux in the maxima of some powerful flares. This circumstance constrained the event selection as well since an event should be sufficiently powerful in order to be registered in more than eight X-ray channels of HEND. For the time instances remained after such a selection the spatial orientation of both Mars Odyssey SC and HEND device mounted on it was determined with the use of the SPICE package \cite{SPICE} in order to determine the detectors illumination conditions.

In each event used in the cross-calibration procedure one and the same time interval was selected in the data of both SC. Each interval was 20 s long (equal to the time cadence of HEND spectral data). For a number of sufficiently long duration flares several time intervals were selected. Using the HESSI and OSPEX packages from SolarSoftWare \cite{Smith02} the absolute radiation fluxes (photon cm$^{-2}$ s$^{-1}$) were obtained in the selected time intervals in the energy ranges corresponding to the HEND energy channels. These fluxes were reduced to the distance to Mars. The values obtained were correlated with the count rates in the corresponding HEND channels. Before the correlation the background from RHESSI and HEND data was subtracted.

\begin{figure}
\includegraphics[scale=0.37]{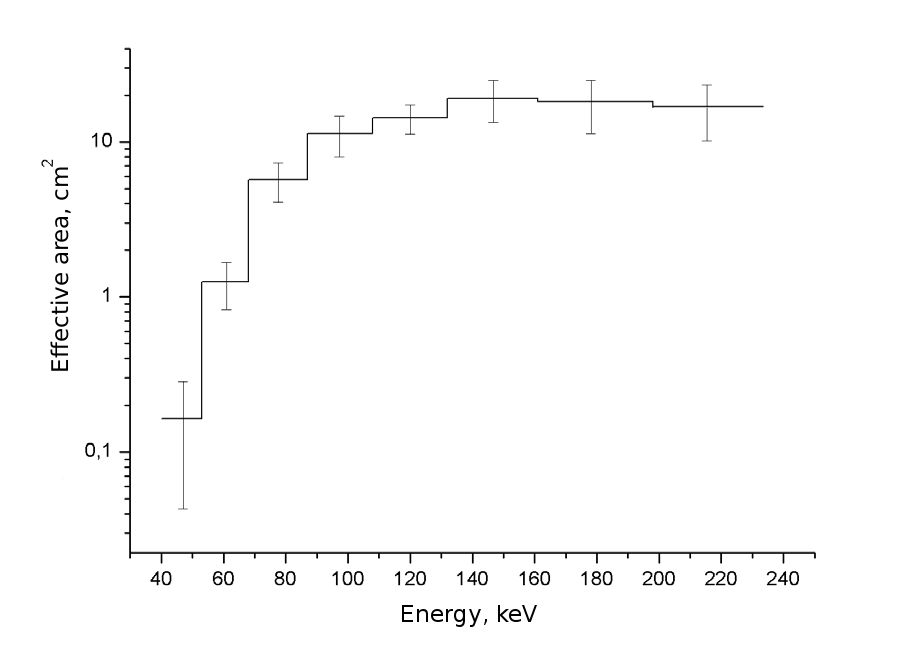}
\caption{\label{effective}The dependence of the outer HEND detector sensitivity on the energy for 1-8 energy channels (the effective area). The cross-correlation with RHESSI was used.}
\end{figure}

Thus, in each energy channel a set of calibrating coefficients $k_i$ was obtained, the number of which (14) was equal to the number of the time intervals selected in 3 flares. Mean values of $k_i$ coefficients and their standard deviations are given in Fig.~\ref{effective}. Numerical values of the coefficients are equal to the effective area of the HEND detector in each energy channel.

Knowing the values of $k_i$ one can find from the HEND count rates the absolute photon fluxes in the range of each energy channel, say, at the distance to Earth:
\begin{equation}
	F_i = \frac{H_i}{k_i}\left(\frac{r_\mathrm{M}}{r_\mathrm{E}}\right)^2,
\end{equation}
where $H_i$ is the count rate at the selected point of the HEND time profile with the background subtracted (counts per second), $r_M$ and $r_E$ are the distances to Mars Odyssey and to Earth respectively at the time of the event in question.

\begin{figure}
\includegraphics[scale=0.20]{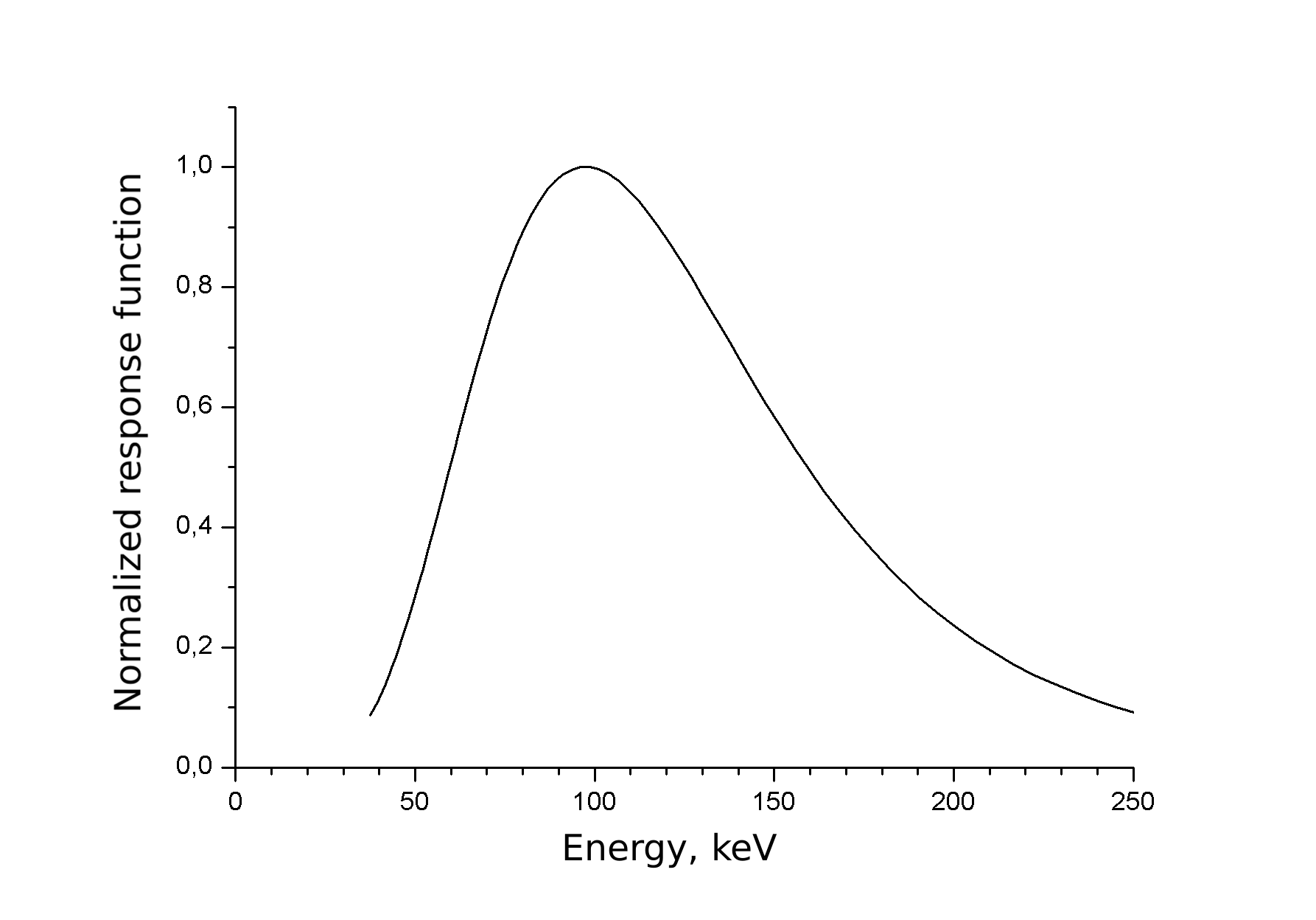}
\caption{\label{response}The curve of the response of the outer HEND detector to the radiation with the power law spectrum with the slope $\gamma=3$. The cross-correlation with RHESSI was used.}
\end{figure}

The simplest way to derive a differential spectrum is to divide $F_i$ by the width of each channel. Usually in the energy range where HEND operates a power law is used to fit the spectrum: $I(E) = I_0E^{-\gamma}$. The parameters $I_0$ and $\gamma$ and their errors were determined via minimization of the weighted sum of square residuals between the model values of the absolute fluxes $F^\mathrm{mod}_i$ and those observed $F_i$. As a lower boundary of the differential spectrum we previously adopted the energy of 50 keV. The accuracy of the spectrum parameters determination depends on the scatter of the $k_i$ value. The absolute value of the root mean square deviation of $k_i$  at 100 keV equals 3 cm$^2$  which corresponds to the relative error of 20\% which keeps in the range from 50 to 250 keV.

A set of programs was written in Python which allow to construct automatically the differential radiation spectrum given the HEND data. The input data are the date and time, time intervals over which the background is estimated and the degree of the polynomial used to approximate the background.

Since in the catalog we give the count rates summed over all the energy channels, it is necessary to show how one can roughly estimate the absolute flux at each point of such a profile. The response function determines the response value of the detector as a function of the energy of the registered radiation. In Fig.~\ref{response} we plot the response function of the outer HEND detector normalized to the maximum value for the radiation with the power law spectrum with the slope $\gamma = 3$. Such a slope is typical for solar flares of M--X classes in the HEND operating range. The effective energy, i.e. the energy of the highest count rate for such a slope equals approximately 100 keV, the half-width of the response function is also about 100 keV. For the spectral slopes $\gamma = 1.5$ and 4.5 these values are equal to 127 and 74 keV for the effective energy and 133 and 77 keV for the half-width, respectively. For such estimations one can use the effective area values plotted in Fig.~\ref{response} at the effective energies indicated.


\end{document}

%% file: table866.dat
      *1 & 12.05.2001 & 23:35:43 &  31.2 &	M3.0	&	9455	&	1.07 $\pm$ 1.28 & 	1.21$\pm$0.93 & 	2.08$\pm$1.63 &  	140 & 	470.8 &		0\\
      2 & 20.05.2001 & 06:02:24 &  39.6 &	M6.4	&	9455	&	18.29 $\pm$ 2.80 & 	17.18$\pm$1.87 & 	3.32$\pm$0.20 &  	140 & 	2863.5 &	0\\
      *3 & 04.06.2001 & 08:08:01 &  58.4 &	M2.5	&	9488	&	2.63 $\pm$ 1.59 & 	2.21$\pm$1.08 & 	3.52$\pm$1.30 &  	40 & 	183.9 &		0\\
      4 & 05.06.2001 & 04:46:08 &  59.5 &	M2.4	&	9488	&	7.53 $\pm$ 2.04 & 	6.77$\pm$1.37 & 	3.35$\pm$0.44 &  	140 & 	1882.9 & 	0\\
      5 & 19.07.2001 & 09:57:39 & 116.5 &	M1.8	&	9537	&	7.63 $\pm$ 1.97 & 	6.38$\pm$1.11 & 	2.70$\pm$0.26 &  	120 & 	1017.8 &	0\\
      6 & 25.08.2001 & 16:30:48 & 157.5 &	X5.3	&	9591	&	256.38 $\pm$ 15.20 & 	327.73$\pm$11.98 & 	2.56$\pm$0.05 &  	280 & 	48570.7 &0\\
      7 & 30.08.2001 & 20:35:35 & 162.2 &	M3.0	&	9601	&	31.26 $\pm$ 5.14 & 	32.89$\pm$3.47 & 	3.65$\pm$0.21 &  	40 & 	2106.6 & 	0\\
      8 & 16.07.2002 & 15:50:48 & 313.5 &	-		&	-		&	78.52 $\pm$ 6.19 & 	84.81$\pm$4.72 & 	3.09$\pm$0.10 &  	1100 & 	32614.0 &	0\\
      9 & 14.10.2002 & 14:26:30 & 332.3 &	-		&	-		&	26.94 $\pm$ 3.57 & 	30.09$\pm$2.49 & 	2.48$\pm$0.12 &  	300 & 	9044.6 & 	1\\
     10 & 25.10.2002 & 09:37:31 & 332.3 &	-		&	-		&	105.17 $\pm$ 7.46 & 	122.61$\pm$5.72 & 	2.83$\pm$0.07 &  	1200 & 	72928.7 &0\\
     11 & 26.10.2002 & 07:06:16 & 332.3 &	-		&	-		&	36.51 $\pm$ 4.02 & 	43.29$\pm$2.93 & 	2.40$\pm$0.10 &  	1180 & 	25729.4 & 	1\\
     12 & 27.10.2002 & 22:53:01 & 332.2 &	-		&	-		&	284.74 $\pm$ 15.92 & 	337.70$\pm$11.37 & 	2.31$\pm$0.04 &  	1060 & 	176017.7 & 0\\
     13 & 27.05.2003 & 23:05:18 & 212.7 &	X1.3	&	10365	&	37.09 $\pm$ 4.84 & 	40.43$\pm$3.54 & 	3.11$\pm$0.16 &  	460 & 	16170.4 & 	0\\
     14 & 28.05.2003 & 00:26:20 & 212.7 &	X3.6	&	10365	&	90.33 $\pm$ 7.32 & 	108.34$\pm$5.89 & 	3.10$\pm$0.10 &  	680 & 	80096.9 & 	0\\
     15 & 29.05.2003 & 01:04:24 & 212.0 &	X1.2	&	10365	&	36.58 $\pm$ 4.87 & 	38.22$\pm$3.67 & 	2.94$\pm$0.20 &  	440 & 	10184.4 & 	0\\
     16 & 31.05.2003 & 02:22:33 & 210.7 &	M9.3	&	10365	&	123.59 $\pm$ 9.60 & 	141.45$\pm$7.56 & 	3.06$\pm$0.09 &  	580 & 	36119.0 & 0\\
     17 & 17.06.2003 & 22:53:33 & 200.6 &	M6.8	&	10386	&	48.99 $\pm$ 5.32 & 	55.52$\pm$3.83 & 	2.59$\pm$0.11 &  	480 & 	18640.0 & 	0\\
     18 & 21.10.2003 & 03:44:00 & 201.7 &	-		&	10486	&	23.15 $\pm$ 4.13 & 	23.57$\pm$3.09 & 	3.49$\pm$0.32 &  	420 & 	6852.6 & 	0\\
     19 & 21.10.2003 & 18:44:53 & 202.1 &	-		&	10486	&	22.06 $\pm$ 2.96 & 	21.65$\pm$2.44 & 	4.55$\pm$0.53 &  	220 & 	11785.0 & 	0\\
     20 & 22.10.2003 & 03:30:01 & 202.2 &	M3.7	&	10484	&	14.23 $\pm$ 3.92 & 	14.34$\pm$2.78 & 	2.62$\pm$0.40 &  	60 & 	559.3 & 	0\\
     21 & 23.10.2003 & 08:25:19 & 202.8 &	X5.4	&	10486	&	95.08 $\pm$ 6.90 & 	104.99$\pm$5.45 & 	3.25$\pm$0.10 &  	480 & 	35415.6 & 	0\\
     22 & 24.10.2003 & 02:48:36 & 203.2 &	M7.6	&	10486	&	32.57 $\pm$ 5.01 & 	34.77$\pm$3.72 & 	3.01$\pm$0.21 &  	900 & 	24151.3 & 	0\\
     23 & 25.10.2003 & 10:29:30 & 203.9 &	M1.5	&	10484	&	24.19 $\pm$ 4.48 & 	24.67$\pm$3.17 & 	3.15$\pm$0.27 &  	140 & 	2126.1 & 	0\\
     24 & 26.10.2003 & 06:14:26 & 204.3 &	X1.2	&	10486	&	29.28 $\pm$ 4.72 & 	29.73$\pm$3.59 & 	3.32$\pm$0.30 &  	520 & 	8755.2 & 	0\\
     25 & 26.10.2003 & 07:23:15 & 204.3 &	-		&	-		&	35.97 $\pm$ 3.86 & 	36.98$\pm$3.31 & 	4.32$\pm$0.40 &  	160 & 	12207.5 & 	1\\
     26 & 26.10.2003 & 07:29:50 & 204.3 &	X1.2	&	10486	&	16.70 $\pm$ 2.64 & 	19.67$\pm$2.01 & 	2.51$\pm$0.18 &  	1140 & 	39978.6 & 	1\\
     27 & 28.10.2003 & 11:03:34 & 205.5 &	X17.2	&	10486	&	359.42 $\pm$ 19.38 & 	462.05$\pm$15.75 & 	2.68$\pm$0.04 &  	1600 & 	686182.9 & 0\\
     28 & 17.11.2003 & 09:01:44 & 216.8 &	M4.2	&	10501	&	34.34 $\pm$ 3.77 & 	33.57$\pm$2.81 & 	3.35$\pm$0.20 &  	200 & 	4858.0 & 	0\\
     29 & 06.01.2004 & 06:22:29 & 248.4 &	M5.8	&	10537	&	31.49 $\pm$ 3.61 & 	35.19$\pm$2.73 & 	2.86$\pm$0.14 &  	160 & 	4915.4 & 	0\\
     30 & 07.01.2004 & 03:56:29 & 249.0 &	M4.5	&	10537	&	24.31 $\pm$ 3.14 & 	27.16$\pm$2.43 & 	2.85$\pm$0.18 &  	320 & 	9253.0 & 	0\\
     31 & 07.01.2004 & 10:22:35 & 249.1 &	M8.3	&	10537	&	15.84 $\pm$ 3.00 & 	16.48$\pm$2.31 & 	3.17$\pm$0.34 &  	380 & 	5444.4 & 	0\\
     32 & 04.09.2004 & 05:44:21 & 326.4 &	-		&	-		&	143.81 $\pm$ 10.21 & 	161.80$\pm$8.15 & 	3.21$\pm$0.09 &  	220 & 	20324.2 & 0\\
     33 & 25.11.2004 & 15:41:45 & 310.5 &	-		&	-		&	23.53 $\pm$ 4.57 & 	27.57$\pm$3.10 & 	2.10$\pm$0.16 &  	60 & 	1583.0 & 	0\\
     34 & 13.05.2005 & 16:41:58 & 199.5 &	M8.0	&	10759	&	11.48 $\pm$ 3.04 & 	11.17$\pm$1.95 & 	2.89$\pm$0.32 &  	560 & 	13409.8 & 	0\\
     35 & 14.07.2005 & 07:22:55 & 182.1 &	M9.1	&	10786	&	200.09 $\pm$ 14.66 & 	224.60$\pm$11.62 & 	3.26$\pm$0.10 &  	120 & 	13964.9 & 0\\
     36 & 14.07.2005 & 10:33:49 & 182.1 &	X1.2	&	10786	&	40.44 $\pm$ 4.09 & 	44.58$\pm$3.10 & 	2.93$\pm$0.13 &  	2300 & 	66856.8 & 	0\\
     37 & 12.02.2010 & 11:25:58 & 333.4 &	M8.3	&	11046	&	27.09 $\pm$ 5.08 & 	33.14$\pm$3.94 & 	2.66$\pm$0.22 &  	60 & 	1156.9 & 	0\\
     38 & 13.02.2010 & 12:43:18 & 333.5 &	C9.6	&	11048	&	2.86 $\pm$ 1.79 & 	4.09$\pm$1.19 & 	1.67$\pm$0.42 &  	40 & 	145.4 & 	0\\
     39 & 03.11.2010 & 12:13:53 & 242.3 &	C4.9	&	11121	&	2.75 $\pm$ 1.72 & 	2.80$\pm$1.13 & 	2.00$\pm$0.74 &  	80 & 	579.7 & 	1\\
     40 & 09.08.2011 & 08:01:42 & 248.7 &	X6.9	&	11263	&	40.02 $\pm$ 5.46 & 	40.09$\pm$3.89 & 	3.12$\pm$0.20 &  	280 & 	7735.9 & 	0\\
     41 & 07.03.2012 & 00:17:58 & 335.1 &	X5.4	&	11429	&	116.16 $\pm$ 7.96 & 	136.55$\pm$6.47 & 	2.99$\pm$0.08 &  	1840 & 	111564.9 & 0\\
     42 & 03.06.2012 & 17:52:56 & 297.1 &	M3.3	&	11496	&	13.34 $\pm$ 2.53 & 	14.90$\pm$1.81 & 	2.48$\pm$0.22 &  	100 & 	2593.2 & 	0\\
     43 & 17.08.2012 & 08:32:33 & 254.2 &	C4.7	&	-		&	16.49 $\pm$ 4.15 & 	15.28$\pm$3.11 & 	3.17$\pm$0.65 &  	60 & 	880.4 & 	0\\
     44 & 18.08.2012 & 03:21:47 & 253.7 &	M1.8	&	11548	&	7.24 $\pm$ 2.01 & 	7.36$\pm$1.48 & 	2.83$\pm$0.45 &  	120 & 	1259.7 & 	1\\
     45 & 20.09.2012 & 14:55:21 & 236.1 &	-		&	-		&	56.03 $\pm$ 5.75 & 	74.31$\pm$4.08 & 	1.94$\pm$0.06 &  	420 & 	16455.4 & 	1\\
     46 & 22.09.2012 & 02:58:53 & 235.4 &	-		&	-		&	30.05 $\pm$ 3.65 & 	32.86$\pm$2.70 & 	2.91$\pm$0.15 &  	160 & 	6078.4 & 	1\\
     47 & 23.10.2012 & 03:15:59 & 221.3 &	X1.8	&	11598	&	7.66 $\pm$ 2.63 & 	9.03$\pm$1.69 & 	1.86$\pm$0.24 &  	100 & 	675.2 & 	1\\
     48 & 28.10.2013 & 01:59:44 & 324.6 &	X1.0	&	11875	&	20.98 $\pm$ 16.55 & 	26.12$\pm$5.89 & 	2.98$\pm$0.39 &  	0 & 	528.7 & 0\\
     49 & 28.10.2013 & 04:37:56 & 324.7 &	M5.1	&	11875	&	6.15 $\pm$ 2.18 & 	5.37$\pm$1.40 & 	1.99$\pm$0.52 &  	100 & 	971.1 & 	0\\
     50 & 28.10.2013 & 20:54:35 & 325.0 &	M1.5	&	11875	&	15.51 $\pm$ 2.69 & 	15.14$\pm$2.01 & 	3.37$\pm$0.35 &  	160 & 	1694.8 & 	0\\
     51 & 19.11.2013 & 10:21:37 & 333.3 &	X1.0	&	11893	&	54.53 $\pm$ 4.96 & 	62.29$\pm$4.00 & 	2.96$\pm$0.13 &  	280 & 	12542.2 & 	0\\
     52 & 25.02.2014 & 00:45:06 & 330.5 &	X4.9	&	11990	&	11.03 $\pm$ 2.59 & 	13.84$\pm$2.06 & 	2.70$\pm$0.29 &  	620 & 	9422.0 & 	0\\
     53 & 10.06.2014 & 11:40:03 & 268.7 &	X2.2	&	12087	&	75.46 $\pm$ 6.21 & 	92.19$\pm$4.71 & 	2.58$\pm$0.07 &  	140 & 	10070.1 & 	0\\
     54 & 10.06.2014 & 12:47:33 & 268.6 &	X1.5	&	12087	&	45.73 $\pm$ 4.73 & 	52.61$\pm$3.68 & 	2.85$\pm$0.13 &  	720 & 	22756.5 & 	0\\
     55 & 11.06.2014 & 09:04:40 & 268.1 &	X1.0	&	12087	&	32.63 $\pm$ 4.54 & 	32.59$\pm$3.09 & 	2.77$\pm$0.17 &  	140 & 	3740.6 & 	0\\
     56 & 22.10.2014 & 01:39:17 & 201.8 &	M8.7	&	12192	&	3.65 $\pm$ 1.77 & 	3.67$\pm$1.26 & 	2.35$\pm$0.73 &  	100 & 	590.0 & 	1\\
     57 & 22.10.2014 & 01:58:16 & 201.8 &	-		&	-		&	4.29 $\pm$ 1.68 & 	6.33$\pm$1.07 & 	1.64$\pm$0.18 &  	60 & 	459.6 & 	1\\
     58 & 22.10.2014 & 14:06:00 & 201.7 &	X1.6	&	12192	&	46.07 $\pm$ 4.58 & 	47.02$\pm$3.47 & 	3.77$\pm$0.18 &  	240 & 	11471.3 & 	0\\
     59 & 05.05.2015 & 22:08:19 & 246.0 &	X2.7	&	12339	&	67.35 $\pm$ 9.09 & 	77.33$\pm$6.30 & 	2.50$\pm$0.10 &  	80 & 	6430.9 & 	0\\
     60 & 04.11.2015 & 03:23:56 & 335.7 &	M1.9	&	12445	&	15.91 $\pm$ 2.72 & 	16.69$\pm$1.98 & 	2.68$\pm$0.23 &  	140 & 	2228.7 & 	0\\

%% file: Livshits_et_al_2017_aps.bbl
\thebibliography{99}
\bibitem{Boynton04}
W.V.~Boynton, W.C.~Feldman, I.G.~Mitrofanov  L.G.~Evans et al., Space Sci. Rev. \textbf{110}, 37 (2004).

\bibitem{Mitrofanov03}
I.G.~Mitrofanov, M.T.~Zuber, M.L.~Litvak, W.V.~Boynton et al., Science \textbf{300}, 2081 (2003).

\bibitem{Sanin04}
A.~Sanin, I.~Mitrofanov, M.~Litvak, A.~Kozyrev et al., Astron. Soc. Pacific Conf. Ser. \textbf{312}, 134 (2004).

\bibitem{Livshits05}
M.A. Livshits, V.A. Chernetskii, I.G. Mitrofanov, A.S. Kozyrev et al., Astron. Rep. \textbf{49}, 916 (2005).

\bibitem{Kashapova08}
L.K. Kashapova, M.A. Livshits, Astronomy Reports. \textbf{52}, 1015 (2008).

\bibitem{Livshits11}
M.A. Livshits, D.V. Golovin, L.K. Kashapova, I.G. Mitrofanov, A.S. Kozyrev et al., Astron. Rep. \textbf{55}, 551 (2011).

\bibitem{Vybornov12}
V.I. Vybornov, M.A. Livshits, L.K. Kashapova, I.G. Mitrofanov et al., Astron. Rep. \textbf{56}, 805 (2012).

\bibitem{Suzaku}
K.~Mitsuda, M.~Bautz, H.~Inoue, R.L.~Kelley et al., Publ. Astron. Soc. Japan \textbf{59}, 1 (2007).

\bibitem{Agostinelli03}
S.~Agostinelli, J.~Allison, K.~Amako, J.~Apostolakis et al., Nuclear Instruments and Methods in Physics Research \textbf{A 506}, 250 (2003).

\bibitem{mcnpx}
R. Lemrani, M. Robinson, V.A. Kudryavtsev, M. De Jesus, G. Gerbier, N.J.C. Spooner. Nuclear Instruments and Methods in Physics Research \textbf{A 560}, 454 (2006).

\bibitem{Lin02}
R.P.~Lin, B.R.~Dennis, G.J.~Hurford, D.M.~Smith et al., Solar Phys. \textbf{210}, 3 (2002).

\bibitem{Smith02}
D.M.~Smith, R.P.~Lin, P.~Turin, D.W.~Curtis et al., Solar Phys. \textbf{210}, 33 (2002).

\bibitem{IZMIRAN}
A. Belov, H. Garcia, V. Kurt, H. Mavromichalaki, M. Gerontidou. Solar Phys. \textbf{229}, 135 (2005).

\bibitem{SPICE}
C.H. Acton. Planetary and Space Science. \textbf{44}, 65 (1996).

\bibitem{Livshits_KI}
M.A. Livshits, A.V. Belov, A.I. Shakhovskaya, E.A. Eroshenko, A.R. Osokin, L.K. Kashapova. Cosmic Research. \textbf{51}, 326 (2013).

\bibitem{Suzaku_catalog}
A.~Endo, T.~Minoshima, K.~Morigami, M.~Suzuki et al., Publ. Astron. Soc. Japan \textbf{62}, 1341 (2010).
